# Modeling the variability of memristive devices with hexagonal boron nitride as dielectric

Juan B. Roldan*, David Maldonado, C. Aguilera-Pedregosa, F. J. Alonso, Yiping Xiao, Yaqing Shen, Wenwen Zheng, Yue Yuan, Mario Lanza* *Member, IEEE*

***Abstract*— Variability in memristive devices based on h-BN dielectrics is studied in depth. Different numerical techniques to extract the reset voltage are described and the corresponding cycle-to-cycle variability is characterized by means of the coefficient of variance. The charge-flux domain was employed to develop one of the extraction techniques, the calculation of the integrals of current and voltage to obtain the charge and flux allows to minimize the effects of electric noise and the inherent stochasticity of resistive switching on the measurement data. A model to reproduce charge versus flux curves has been successfully employed. The device variability is also described by means of the time series analysis to assess the "memory" effect along a resistive switching series. Finally, we analyzed I-V curves under ramped voltage stress utilizing a simulator based on circuit breakers, the formation and rupture of the percolation paths that constitute the conductive nanofilaments is studied to describe the set and reset processes behind the resistive switching operation.**

## I. Introduction

MEMRISTIVE devices have shown promising features for applications linked to non-volatile memories, neuromorphic computing, mobile communication, and data encryption [1-7]. Memristive devices made of two-dimensional (2D) layered materials exhibit multiple advantages compared to state-of-the-art memristive devices made of phase-change, metal-oxide or magnetic materials [3]. Among them, the most remarkable are: high transparency (i.e., ~98% light transmittance) [8], high flexibility (i.e., stable operation under bending radius up to ~1.2 cm) [9], coexistence of volatile threshold-type and non-volatile bipolar switching [10], better controllability of potentiation and relaxation processes [7], ultra-low switching energy (down to ~8.8 zJ) [1], and high thermal stability up to 300 ºC [11]. Among all 2D materials for memristive devices, multilayer hexagonal boron nitride (h-BN) has exhibited the best yield and performance as resistive switching medium due to its high bandgap (~6 eV) — which is necessary to block the leakage current — and high mechanical features (compared to monolayers) that avoid fracture during transfer (resulting in a higher yield).

However, one of the main obstacles towards the construction of commercial integrated circuits containing 2D materials-based devices (of any kind, not only memristive devices) is to increase the yield and reliability, and reduce the device-to-device variability [12]. While the first two items are simple to quantify indicating the yield-pass criteria [13] and conducting endurance test [14]; the community working in this field is much less familiar with the methodologies to be employed to quantify variability. In the context of memory applications for memristive devices, two different non-volatile resistance states, usually referred to as low resistance state (LRS) and high resistance states (HRS), can be induced by applying electrical stresses. The determination of the starting and ending points of the HRS-to-LRS (i.e., set) and LRS-to-HRS (i.e., reset) transitions and the corresponding resistance states that come up after them is essential to understand the variability of memristive devices, which at the same time is key for the construction of more elaborated memristive circuits.

Compact models have to be developed to enable the fabrication and simulation infrastructure needed at the industrial level, taking into consideration the device hysteretic behaviour, variability effects, both at device-to-device and cycle-to-cycle levels, and thermal effects (among others). In the field of metal-oxide memristive devices, a high number of manuscripts devoted to compact modeling have been published in the last few years [15-22]; however, in the

This work has been supported by the Ministry of Science and Technology of China (2018YFE0100800), the National Natural Science Foundation of China (grant no. 61874075), the Consejería de Conocimiento, Investigación y Universidad, Junta de Andalucía (Spain) and European Regional Development Fund (ERDF) under projects A-TIC-117-UGR18, A-FQM-66-UGR20, A-FQM-345-UGR18, B-TIC-624-UGR20 and IE2017-5414, the grant PGC2018-098860-B-I00 supported by MCIU/AEI/FEDER, and the "Maria de Maeztu" Excellence Unit IMAG, reference CEX2020-001105-M, funded by MCIN/AEI/10.13039/501100011033/. M.L. acknowledges generous support from the King Abdullah University of Science and Technology.
(corresponding authors: mario.lanza@kaust.edu.sa, jroldan@ugr.es)

David Maldonado, C. Aguilera-Pedregosa, and Prof. Juan B. Roldan are with the Departamento de Electrónica y Tecnología de Computadores, Facultad de Ciencias, Universidad de Granada, Avd. Fuentenueva s/n, 18071 Granada, Spain

Francisco J. Alonso is with the Departamento de Estadística e Investigación Operativa e Instituto de Matemáticas IMAG, Universidad de Granada, Facultad de Ciencias, Avd. Fuentenueva s/n, 18071 Granada, Spain.

Yiping Xiao, Yaqing Shen and Wenwen Zheng are with the Institute of Functional Nano & Soft Materials (FUNSOM), Collaborative Innovation Center of Suzhou Nano Science &Technology, Soochow University, 199 Ren-Ai Road, Suzhou 215123, China.

Prof. Mario Lanza and Yue Yuan are with the Physical Sciences and Engineering Division, King Abdullah University of Science and Technology (KAUST), Thuwal 23955-6900, Saudi Arabia.

context of devices with 2D dielectrics for memristive devices there are very few articles that tackle modeling issues [23, 24]. More specifically, even in the domain of widely-investigated metal-oxide memristors very few works discuss reliable parameter extraction methods [25, 26]. On this subject, the lack of literature is aggravated by the numerical difficulties to deal with memristive device experimental data [25, 27, 28, 29].

In this article, we focus on the experimental characterization and memristive effects in Au/Ti/h-BN/Au devices. Different extraction techniques for reset and set voltages are presented [26, 30]. We also account for the resistive switching description along a series a successive set/reset cycles by using the time series analysis and modeling tools. Moreover, the conductivity is analyzed describing the formation and disruption of conductive nanofilaments using a simulation tool based on circuit breakers.

## II. DEVICE FABRICATION AND MEASUREMENT SET-UP

The memristive device stack (from top to bottom) consists of 40 nm Au, 10 nm Ti, 6 nm h-BN, 40 nm Au, and it was fabricated on a Si wafer with a 300 nm thick $SiO_2$ layer on top. The first step consisted of the bottom electrodes deposition by means of a square pad of 100 µm × 100 µm connected to a metallic wire with a width of 5 µm. A progressive reduction of the width from 100 µm to 50 µm allowed the transition between them. The electrodes were deposited by photolithography, electron beam evaporation and lift-off. A mask aligner MJB4 from SUSS MicroTech and an electron beam evaporator PVD 75 from Kurt Lesker were employed. After the bottom electrode deposition, a sheet of ~18-layers-thick h-BN, grown independently on a Cu substrate by chemical vapor deposition method at the laboratories of Graphene Supermarket, was transferred on the bottom electrodes. Finally, top electrodes with the same shape than the bottom ones but rotated 90 degrees are deposited, forming a cross-point Au/Ti/h-BN/Au region with lateral size of 5 µm × 5 µm. Between the bottom Au electrode and the $SiO_2$ substrate a 10-nm-thick layer of Ti was deposited (by electron beam evaporation) to favor adhesion. The device fabrication steps are depicted in Fig. 1. The electrical measurements were carried out using a Keysight B1500A semiconductor parameter analyzer connected to a probe station (Karlsuss PSM6) provided with a B1511B medium power source measurement unit (MPSMU) module for quasi-static ramped voltage stress (RVS).

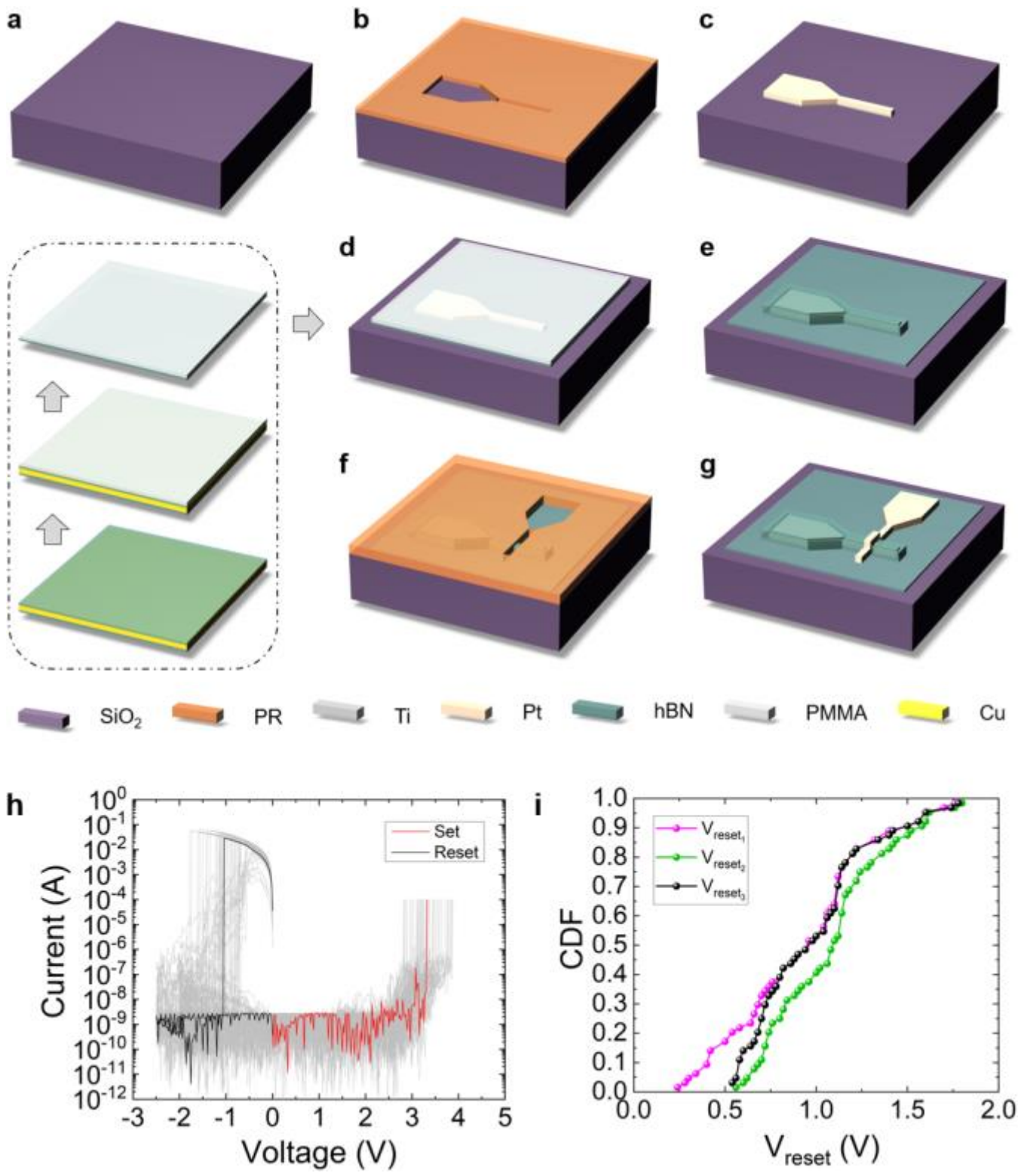

Fig. 1. Resistive switching behavior of Au/Ti/h-BN/Au/Ti memristive devices. (a) to (g) Fabrication steps of the memristive devices. The dielectric consists of a 6 nm thick layer of h-BN. (h) Experimental current versus applied voltage in a long RS series for our devices (the compliance current used was $I_{CC}$=0.1 mA). (i) Cumulative distribution function for the three different reset voltages calculated: $V_{reset1}$ (maximum current value), $V_{reset2}$ (determined at the point corresponding to $Q_{reset}$), $V_{reset3}$ (minimum current derivative, as shown in Fig. 2(b).

## III. SIMULATOR DESCRIPTION

We employ a flexible simulation tool based on circuit breakers (CBs). This computational tool allows the step-by-step description of the formation and rupture processes of conductive nanofilaments in a 2D domain. The simulator includes CBs with up to four conductance levels, it accounts for quantum effects in charge conduction by means of the quantum point contact (QPC) model, as well as for the series resistance and dielectrics with several layers [31]. A 2D network represents the conduction structure of the dielectric, including the possibility of percolation paths formation that represent metallic conductive nanofilaments that short the electrodes after a successful set process. A device with a pristine dielectric is established to initiate the simulation, where some CBs are randomly chosen to be in LRS to reproduce the stochastic behavior of a real memristive device. If CBs with three or four levels are chosen, the different resistance values are employed for the random initialization of the dielectric. Joule heating is considered and a thermal-driven switching rule for the CB is also implemented.

## IV. RESULTS AND DISCUSSION

The devices were measured by alternating positive and negative ramped voltage stresses (RVS) applied to the top electrode while keeping the bottom grounded. We apply long series of more than 60 cycles using a current limitation of 100 μA during the positive RVS, which is necessary to avoid damage of the device during the set transition — the negative RVS is not current-limited and triggers the LRS-to-HRS switch reset. During the first RVS the devices show low conductance (~0.01 μS) and a relatively high dielectric breakdown voltage (ranging between 4 and 8 V for different devices). After that, the devices show non-volatile bipolar RS (see Fig. 1h), with HRS and LRS conductance of ~0.01 μS and ~0.05 S (respectively). The RS operation in these devices was shown [10] to be provoked by the creation and rupture of Ti-based conductive nanofilaments across the h-BN stack, probably at the native defects embedded within the crystalline 2D layered lattice, due to the lower energy for metal penetration at such sites [32].

This switching mechanism presents a certain degree of stochasticity, i.e., the values of the set and reset voltage ($V_{set}$, $V_{reset}$) and the conductance in HRS and LRS ($G_{HRS}$, $G_{LRS}$) can change from one cycle to another [3, 14]. This variability is normally quantified by calculating the coefficient of variance (CV), that can be obtained by dividing the standard deviation (σ) by the mean value (μ) of the distributions of values. Note that there is no clear boundary between what is considered acceptable in terms of variability, as that is an application-specific metric that should be defined within the yield-pass criteria. In general, if the variability of such parameters from cycle to cycle is small (i.e., $C_V$ of $V_{set}$ < 2%), the devices could be employed for information storage [33], computation [34] or transmission [35]; and if the variability is high ($C_V$ of $V_{set}$ >20%), the devices could be employed for data encryption as entropy source of true random number generators [5] or physical unclonable functions [36]. Hence, understanding and quantifying the variability is critical. In a memristive device based on conductive nanofilament formation and disruption, the reset process normally exhibits higher variability than the set, i.e., $V_{reset}$ presents a higher dispersion than the $V_{set}$ and the conductance after the reset ($G_{HRS}$) presents a higher dispersion than the current after the set ($G_{LRS}$) [1, 3, 37]. The reason is that the current limitation employed during the set process homogenizes the size of the conductive nanofilaments. In order to evaluate the worst-case scenario, in the following we focus on the analysis of the variability of $V_{reset}$.

We extract the values of $V_{reset}$ from the I-V curves (Fig. 1(h)) by using three different methods. The first method consists of reading the voltage at which the LRS current ($I_{LRS}$) is maximum in each RVS (see Fig. 2(a)). This method is the simplest and it has been used in multiple studies [26, 38, 39]. In the second method, we use the charge (Q) and the flux (ϕ). The charge can be calculated from the conventional I-V curves as follows:

$$Q(t) = \int_0^t i(t')\,dt' \qquad (1)$$

where i(t) and v(t) are the measured current and voltage. And the flux can be calculated as follows:

$$\phi(t) = \int_0^t v(t')\,dt' \qquad (2)$$

Fig. 2(c) shows the Q-ϕ curves that come out of the I-V curves in Fig. 1(h). We have implemented a new technique for $V_{reset}$ extraction that is related to the determination of $Q_{reset}$ and $\phi_{reset}$; both of them are shown in Fig. 2(c)-2(e). These two parameters are obtained at the point in which a null charge derivative with respect to ϕ is detected; taking into account (1) and (2), this means that the device current drops off to negligible values at the point ($Q_{reset}$, $\phi_{reset}$). This methodology is simple and numerically stable because the first integrals of the current and voltage allow to eliminate any noise in the measurements. The third method calculates $V_{reset}$ corresponding to the value in which the derivate of $I_{LRS}$ is minimum (see Fig. 2(b)). Calculating the derivate of $I_{LRS}$ is challenging because this current changes during reset processes, and also due to the electrical noise; therefore, different numerical techniques can be employed [25, 27, 28, 29].

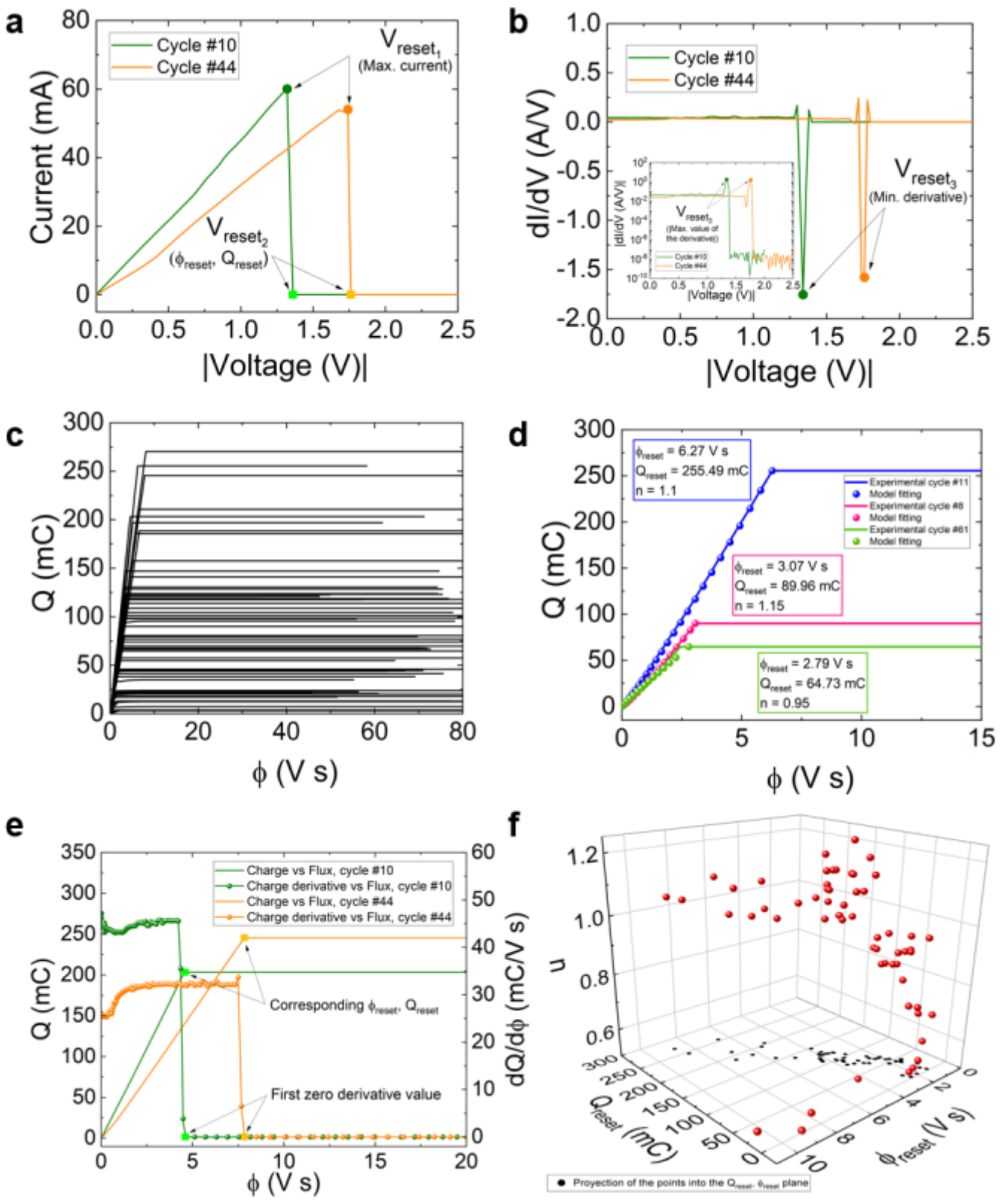

Fig. 2. Resistive switching parameter extraction in Au/Ti/h-BN/Au/Ti memristive devices. (a) Current versus absolute value of the voltage for cycles #10 and #44, where two different numerical techniques to extract the reset voltage are shown. $V_{reset1}$ is determined at the maximum current value and $V_{reset2}$ is established at the first point where the charge derivative with respect to the flux is null. (b) Current derivative versus absolute value of the applied voltage for the two cycles in Figure 2a, an inset is included where the absolute value of this derivative is shown in in logarithmic scale for clarity. This technique is employed to determine $V_{reset3}$. (c) Charge versus flux calculated by means of the experimental I-V curves (obtained with (1) and (2)). (d) Charge versus flux and modeled data calculated through (3). The $Q_{reset}$ and $\phi_{reset}$ are determined as indicated in Fig. 2(a). (e) Charge versus flux for two cycles, #10 and #44, and the corresponding derivative of the charge with respect to the flux, performed to extract the $Q_{reset}$ and $\phi_{reset}$ points at the first null derivative value. (f) The n coefficient has been obtained by minimizing the square mean error of the difference between the experimentally calculated charge and that given in (3) (a semiempirical model introduced in [40]).

In relation to the representation of the I-V data (Fig. 1(b)) in the charge-flux domain (Fig. 2(c)), the modeling of these latter curves is much easier. Equation 3 [40] shows a compact expression that allows the correct fitting of the Q-ϕ curves, some of the fitting constants for our data are given in Fig. 2(d) and Fig. 2(f).

$$Q(\phi) = Q_{reset}\left(\frac{\phi}{\phi_{reset}}\right)^{n} \qquad (3)$$

where ϕ is the flux obtained in (2) and n, a fitting parameter.

It is important to highlight that the modeling of the I-V curves is much more complicated [15-22]; on the contrary, three parameters allow an accurate Q-ϕ curve fit (Figure 2(d)). The n parameters employed ($0.6<n<1.3$) for the curve set shown in Fig. 2(c) are plotted in Fig. 2(f), see that n increases as the values of $Q_{reset}$ and $\phi_{reset}$ rise. The cumulative distribution functions (CDF) of $Q_{reset}$ and $\phi_{reset}$ are shown in Fig. 3.

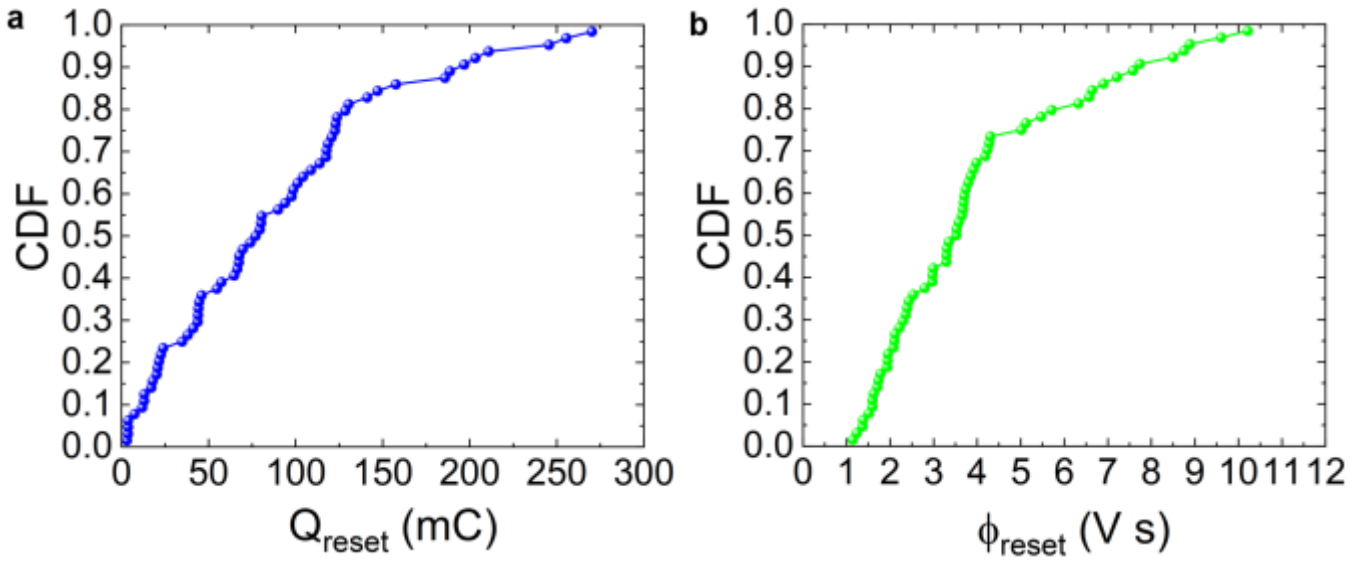

Fig 3. (a) Cumulative distribution functions for the $Q_{reset}$ and (b), $\phi_{reset}$ obtained for the whole RS series making use of (1) and (2) and the numerical procedure explained in Fig. 2(d) and Fig. 2(e).

See that $\phi_{reset}$ shows indirectly the voltage needed to reach the reset point where the conductive nanofilament is destroyed. If the voltage is understood as a RVS signal, $v(t) \approx m \cdot t$ (where m is the slope of the voltage signal), then $\phi(t) \approx 1/2 \cdot m \cdot t^2$, $Q_{reset}$ stands for the charge extracted from the device (assuming that the opposite current injects charge, as it would correspond in the set process) to trigger the reset process.

Fig. 1(c) shows a comparison of the values of $V_{reset}$ determined with the three methods. As expected, the cycle-to-cycle variability depends on the extraction technique. The variability of $V_{reset}$ obtained using method 1 (the maximum current value in the I-V curve) shows the higher dispersion ($C_{V1}$ = 40%), and the dispersion of $V_{reset}$ using methods 2 and 3 (corresponding to the point of $Q_{reset}$ and the minimum of the current derivative, respectively) show closer variabilities ($C_{V2}$ = 28% and $C_{V3}$ = 32%, respectively). In any case, the cycle-to-cycle variability of $V_{reset}$ parameters is similar to that observed in reference metal-oxide-based memristive devices [19]. This representation can be easily analyzed using times series modeling [23] and implemented in Verilog-A to extract the "memory effect" feature along multiple successive cycles for circuit-level simulations. The data have been analyzed and shown in Fig. 4. See that the $V_{set}$ data do not present autocorrelation, i.e. if we pick a datum in the RS series it does not show dependence on the data corresponding to previous cycles; therefore, no time series model can be extracted [15, 41, 42]. This is not the case of $V_{reset}$, the data are autocorrelated (there exist dependence between data of different cycles in the RS series) and a time series model can be extracted (see (4)) following the procedures explained in [15, 27, 41, 42]. In particular, the reset voltage time series model was obtained using the Box-Jenkins methodology [42]. An Autoregressive Integrated Moving Average (known as ARIMA) model [42] was proposed.

$$V_{reset_t} = V_{reset_{t-1}} - 0.8293\,\varepsilon_{t-1} \qquad (4)$$

where $V_{reset_t}$ is the modeled reset voltage in the current cycle of an RS series, and $V_{reset_t-1}$ is the modeled reset voltage lagged one cycle (i.e., the reset voltage value of the previous RS cycles), $\varepsilon_{t-1}$ is the error (the residual in the time series argot) made in the modeling process in the previous cycle; i.e, the experimental minus the modeled values [15, 41, 42].

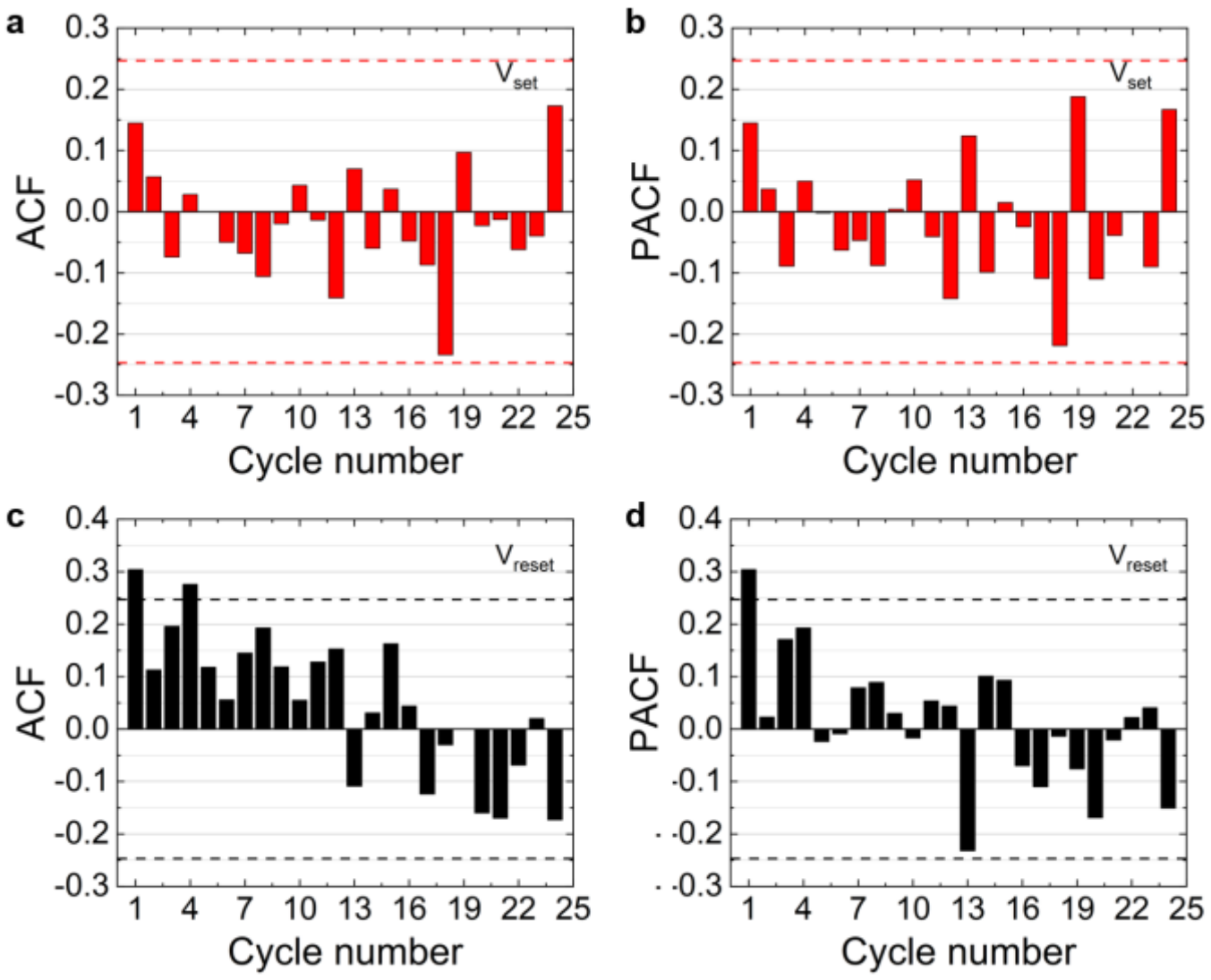

Fig. 4. (a) Autocorrelation Function (ACF) and (b) Partial Autocorrelation Function (PACF) versus cycle lag (distance apart in cycles within a RS series; for a cycle lag 1, the ACF and PACF of consecutive cycles are measured and so on) for the $V_{set}$ series. (c) ACF and (d) PACF versus cycle lag for the $V_{reset}$ series. The ACF and PACF approximate minimum threshold bounds for the devices under study are $\pm 1.96/\sqrt{n}$, where n is the number of cycles in the series shown with dashed lines. A more elaborate calculation was performed for the detailed evaluation performed here [15, 41, 42]. The ACF and PACF data below the threshold bound means non correlated data.

See that the general trend of the data series can be predicted with the model (Fig. 5), as it is expected for a model based on this theory [15, 27]. It is interesting to highlight that the data autocorrelation in this technology is lower than that observed in other conventional resistive memory devices with dielectrics made of transition metal oxides [15]. This fact implies a higher degree of randomness in our devices that could be linked to the different nature of resistive switching processes (the formation of conductive nanofilaments (CNFs) is not as strong, and in the reset processes the CNF remnants do not keep the shape corresponding to previous cycles). Consequently, a greater redistribution of metallic atoms or defects that configurate the CNF is produced, this effect can be behind the giant random conductance fluctuations observed in h-BN devices [43].

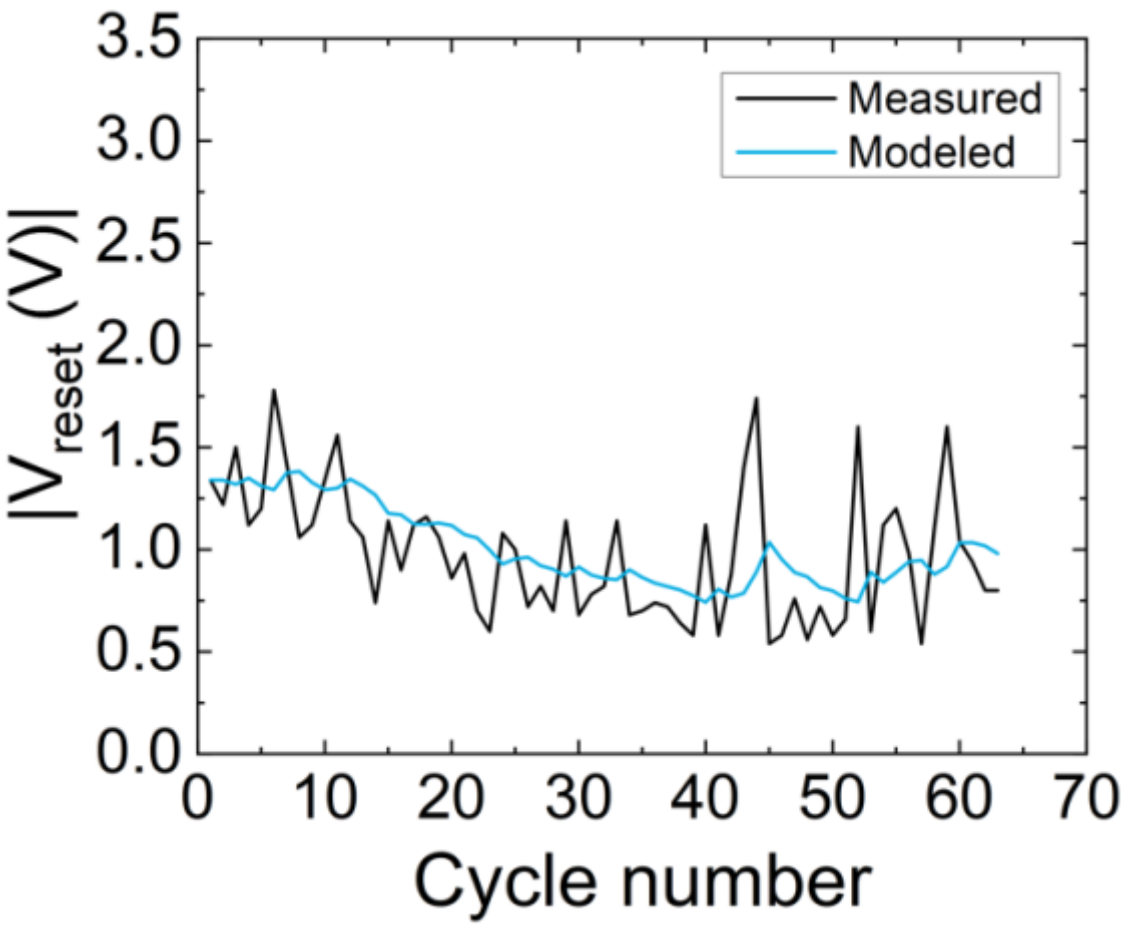

Fig. 5. Absolute value of $V_{reset}$ versus cycle number for the RS series under consideration, the measured values are shown in black and the modeled in blue.

In order to analyze the switching characteristics in a more comprehensive manner, we have made use of a CB simulator [31]. The conduction is described by the formation and disruption of conductive nanofilaments that enable the resistive switching operation. Fig. 6 shows the results of the simulator employed to reproduce experimental I-V curves, both for the reset and set processes.

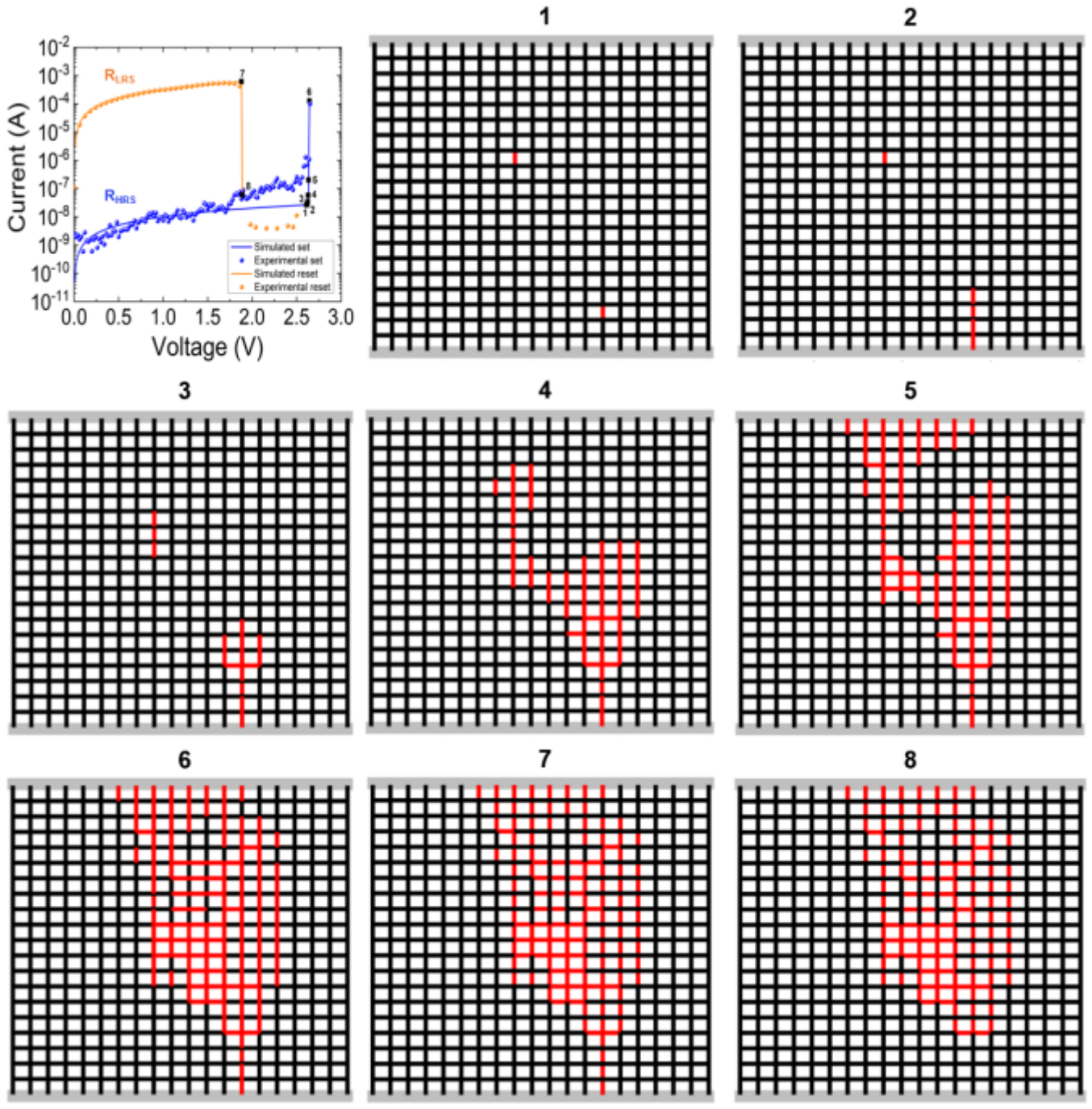

Fig. 6. Circuit breaker simulation of an experimental set and reset cycle. Simulated [31] and experimental current versus voltage curves. The snapshots of the circuit breaker networks linked to the points highlighted along the curves are also given. The values employed in the circuit breaker model shown in Fig. 6 where (for the $R_{off} = 1\times10^8\ \Omega$ and $R_{on} = 300\ \Omega$ and the $V_{off} = 0.182$ V and $V_{on} = 0.17$ V).

The fitting was performed using a simple model to describe the CB conduction properties (Fig. 7).

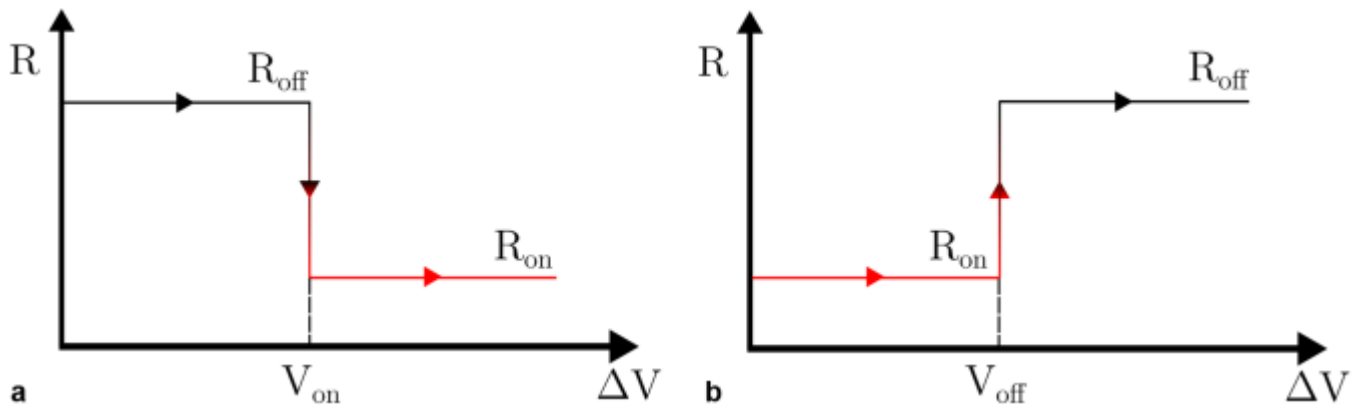

Fig 7. (a) Circuit breaker internal resistance structure for the set (or forming) process and (b), for the reset process. The circuit breaker internal resistance structure is described in two levels, $R_{off} = 1\times10^8\ \Omega$ and $R_{on} = 300\ \Omega$, and the $V_{off} = 0.182$ V and $V_{on} = 0.17$ V.

The two values of the CB resistance are assumed (depending on the voltage in the CB, see the scheme in Fig. 6), consequently, the CNFs are formed with paths made of CBs showing their low resistive component (see Fig. 6, snapshot 6). The CB high resistance component accounts for a defect-free dielectric region and the low resistance component could be associated approximately with the presence of a Ti atom inside the h-BN stack. Assuming a Ti atom radius of 176 pm [44] and a percolation path of atoms in a perfect straight line, we would need around 17 Ti atoms to short the electrodes in a 6 nm dielectric. In this respect, we are considering a 20×20 matrix in our simulations that suffices to model paths of different shapes, even with non-straight sections (wider matrices could be employed for a higher accuracy, which would increase the computing time). See that the curve fitting is reasonably good (Fig. 6), mostly for the reset curve, when the CNF is fully formed. The snapshots of the percolation paths in Fig. 6(1-8) correspond to the points marked in the simulated I-V curves (a voltage map in the CB matrix for the snapshots is shown in Fig. 8). Even though the CNF can have wide sections, the rupture of the highly conductive percolation path in one or several rows allows to reproduce the abrupt current decrease that is observed experimentally. See in the set process that a fast growth of the CNF is obtained with a slight variation of the applied voltage.

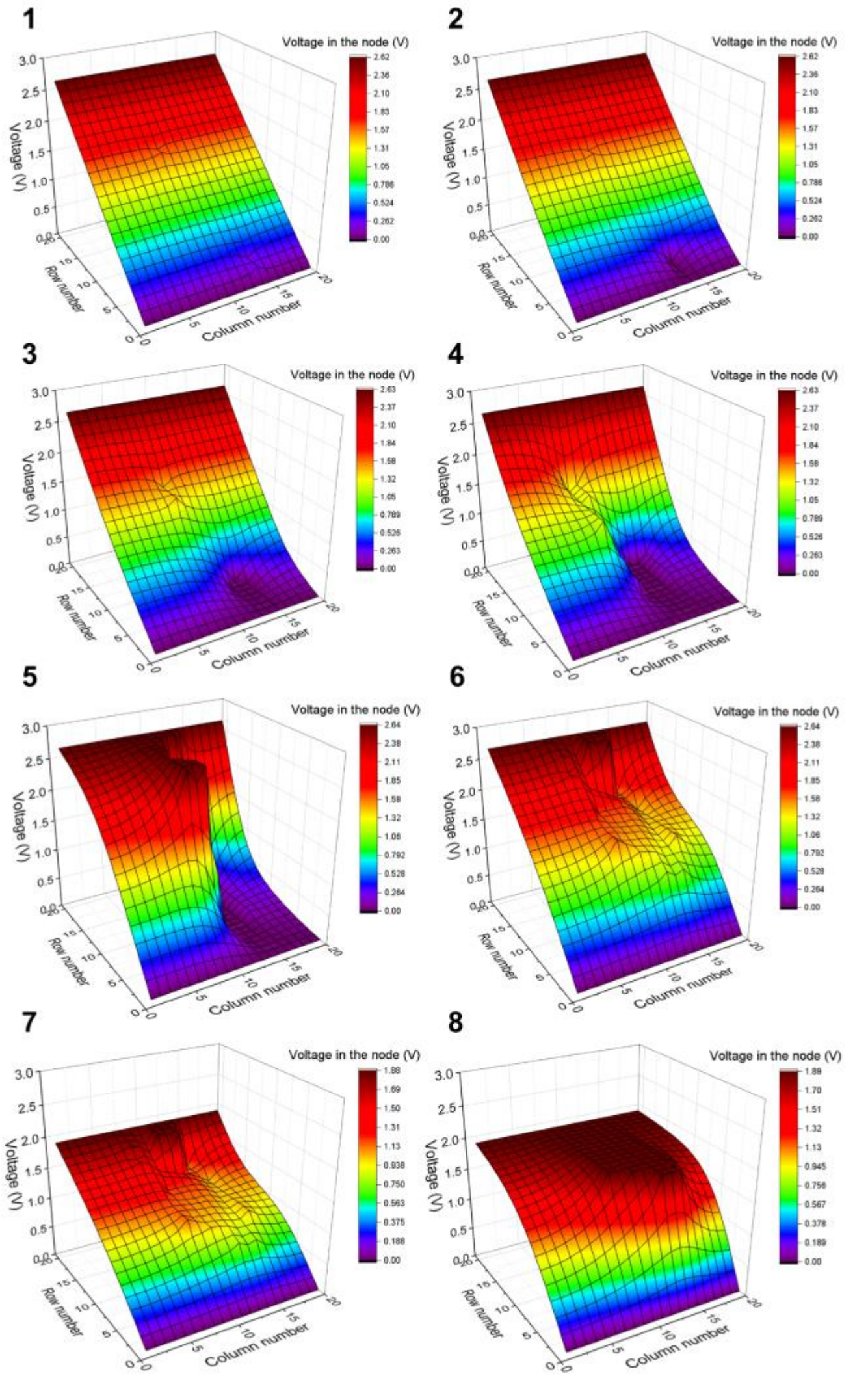

Fig. 8. 3D representation of the voltage in the circuit breaker matrix for the simulation points shown in Figure 6.

## V. CONCLUSIONS

The variability of memristive devices fabricated with h-BN dielectrics and metal electrodes is analyzed with several techniques. The reset voltage was extracted with different numerical techniques, it was shown that the coefficient of variance depends on the procedure employed, it was lower for the reset voltage determined with the current derivative minimum, and at the point where the charge derivative with respect to the flux was null. A model was proposed to describe the device operation in the charge-flux domain, and the fitting parameters were analyzed statistically. “Memory effects” in the reset and set voltages along a resistive switching series were studied using the time series analysis, the actual value of reset voltage is calculated employing previous values of this parameters in the series. The formation and rupture of percolation paths that short the electrodes to allow the switching operation are finally characterized using a simulator based on circuit breakers, where snapshots of a 2D resistive matrix allow to follow the evolution steps of the conductive nanofilaments.